\documentclass[prb, floatfix, nobibnotes, reprint, superscriptaddress]{revtex4-1}
\usepackage[english]{babel}
\usepackage[utf8]{inputenc}
\usepackage[T1]{fontenc}
\usepackage{caption}
\usepackage{amsmath}
\usepackage{mathalpha}
\usepackage{amssymb}
\usepackage{amsfonts}
\usepackage{graphicx}
\usepackage{subcaption}
\usepackage{rotating}
\usepackage{relsize}
\usepackage{bm}
\usepackage{multirow}
\usepackage{bigints}
\usepackage[varg]{txfonts}
\usepackage{bbm}
\usepackage{listings}
\usepackage[hmargin=2.5cm, vmargin=4cm]{geometry}
\graphicspath{{figures/}}
\usepackage{url}
\usepackage{braket}
\usepackage{textcomp}
\usepackage{float}
\usepackage{bbold} 
\usepackage{comment}   
\usepackage{suffix}
\usepackage[dvipsnames]{xcolor} 
\usepackage[normalem]{ulem}

\newcommand{\papertitle}{Alchemical insights into approximately quadratic energies of iso-electronic atoms}

\def\beq{\begin{equation}}
\def\eeq{\end{equation}}
\def\bea{\begin{eqnarray}}
\def\eea{\end{eqnarray}}
\def\brcl{\begin{array}{rcl}}
\def\bccl{\begin{array}{ccl}}
\def\blcl{\begin{array}{lcl}}
\def\err{\end{array}}

\begin{document}

\title{\papertitle}

\author{Simon León Krug}
\affiliation{Machine Learning Group, Technische Universit\"at Berlin, 10587 Berlin, Germany}

\author{O. Anatole von Lilienfeld}
\email{anatole.vonlilienfeld@utoronto.ca}
\affiliation{Machine Learning Group, Technische Universit\"at Berlin, 10587 Berlin, Germany}
\affiliation{Berlin Institute for the Foundations of Learning and Data, 10587 Berlin, Germany}
\affiliation{Chemical Physics Theory Group, Department of Chemistry, University of Toronto, St. George Campus, Toronto, ON, Canada}
\affiliation{Department of Materials Science and Engineering, University of Toronto, St. George Campus, Toronto, ON, Canada}
\affiliation{Vector Institute for Artificial Intelligence, Toronto, ON, Canada}
\affiliation{Department of Physics, University of Toronto, St. George Campus, Toronto, ON, Canada}
\affiliation{Acceleration Consortium, University of Toronto, Toronto, ON, Canada}

\date{\today}

\begin{abstract}
Accurate quantum mechanics based predictions of property trends are so important for materials design and discovery that even inexpensive approximate methods are valuable.
We use the Alchemical Integral Transform~(AIT) to study multi-electron atoms, and to gain a better understanding 
of the approximately quadratic behavior of energy differences between iso-electronic atoms in their nuclear charges.  
Based on this, we arrive at the following simple analytical estimate of 
energy differences between any two iso-electronic atoms, 
$\Delta E \approx -(1 + 2\gamma \sqrt{N_e-1})\,\Delta Z\, \bar{Z}$. 
Here, $\gamma \approx 0.3766 \pm 0.0020$~Ha corresponds to an empirical constant, 
and $N_e$, $\Delta Z$, and $\bar{Z}$ respectively to electron number, and nuclear charge difference and average.
We compare the formula's predictive accuracy using experimental numbers and non-relativistic, numerical results obtained via DFT~(\texttt{pbe0}) for the entire periodic table up to Radon.
A detailed discussion of the atomic Helium-series is included. 
\end{abstract}

\maketitle


\section{Introduction}
\label{sec:introduction}

The electronic quantum many-body problem is without a doubt one of the outstanding challenges of materials design to date. More often than not, only numerical solutions are possible to obtain, but these are associated with complex and costly computations or, in case of modern machine learning~(ML) solutions, subject to trillions of parameters. This offers little intuitive understanding even for systems with few electrons and/or model potentials. Naturally, it is desirable to predict the behavior of many atoms as in molecules and crystals, but even the multi-electron atom, i.e. the periodic table, requires expensive simulation for results below chemical accuracy.

This is where quantum alchemical methods\cite{marzari_1994,putrino_2000,lilienfeld_2009,perez_2011,miranda-quintana_2017,giorgio_2022,shirogawa_2023} provide a resourceful alternative: using few single point calculations, properties of many seemingly unrelated systems can be predicted by exploiting similarities in \textit{relative} changes\cite{beste_2006}. 
These relative approaches often reduce to changes in nuclear composition $\lbrace Z_I \rbrace$ or other parameters. Alchemical transformations, i.e. either alchemical Taylor expansions\cite{von_Rudorff_2020,keith_2020,vonlilienfeld2023_1st_order,balawender_2013}, 
especially up to first\cite{lesiuk_2013,munoz_2017} or second order\cite{rudorff_2020,munoz_2020}, 
or thermodynamic integration (along a transmutation parameter~$\lambda$)\cite{krug_generalAIT} have been used in the context of atoms across the periodic table\cite{balawender_2019} (\textit{vide infra}), 
as well as diatomics\cite{eikey_2022} and polycyclic hydrocarbons\cite{rudorff2020_alchem_chirality}.

In a previous paper\cite{krug_generalAIT}, we derived and discussed a general version of the Alchemical Integral Transform~(AIT); it allowed its user to recover the energy and electron density of a final system~$E_B$ from an iso-electronic initial system's electron density~$\rho_A$ and energy~$E_A$, in $n$ dimensions and for multi-electron systems, if the coordinates of the initial and final Hamiltonian could be expressed as one another by an affine transformation.
Here, we apply AIT to real systems, i.e.~the multi-electron atom. 
We introduce a simple and inexpensive formula for relative energies that even outperforms non-relativistic DFT computations.
We assess the model's numerical performance for the He-atom series, and we investigate its applicability towards the prediction of electron affinities using ionization energies only. 

\begin{figure}
    \centering
    \includegraphics[width=\linewidth]{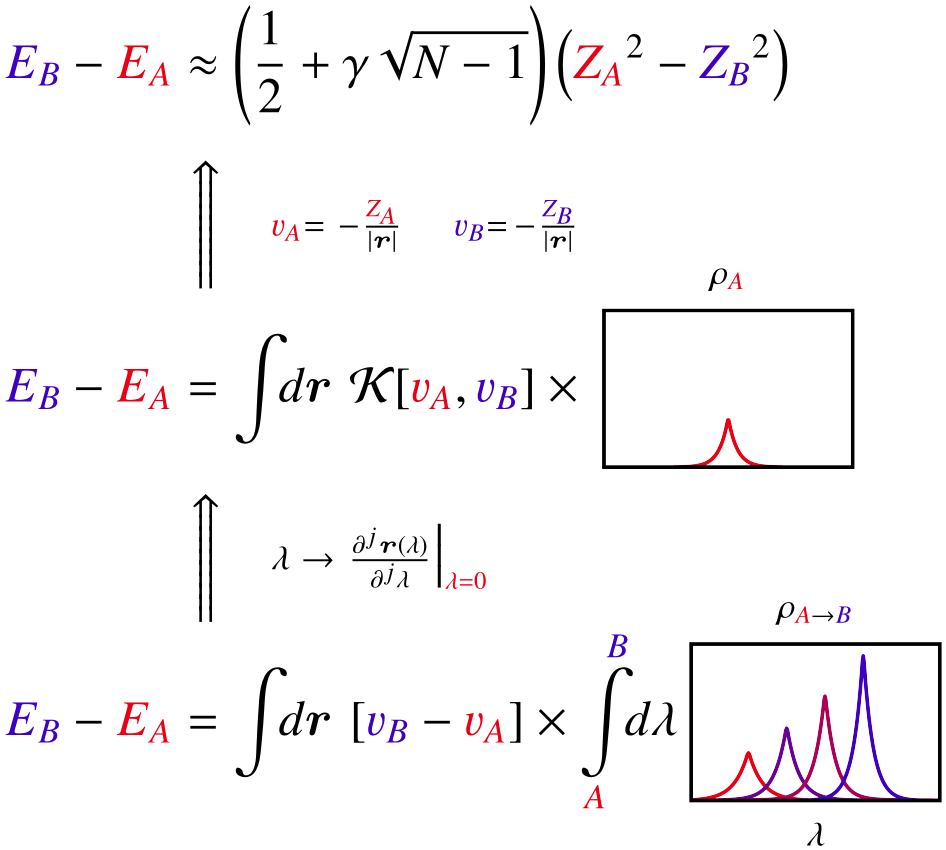}
    \caption{Overview: 
     Approximate energy trends among iso-electronic atoms are revealed by $\mathcal{K}$ for mono-atomic potentials (top).
     Parametrization $\bm{r}(\lambda)$ moves all dependencies of the alchemical path to AIT's kernel $\mathcal{K}[v_B - v_A]$ (mid).
     Depiction of thermodynamic integration between systems $A$ and $B$ (bottom)}
    \label{fig:alchemy_methods}
\end{figure}

\section{Methods}
\label{sec:methods}

The Alchemical Integral Transform (AIT) expresses energy and electron density of a system $B$ using energy and electron density of some iso-electronic reference system $A$. 
This is possible if both systems can be expressed as one another by an affine transformation $A(\lambda) \, \bm{x} + \bm{b}(\lambda)$ of the coordinates of their Hamiltonians, without regard for any normalization \cite{krug_generalAIT}: 
\begin{align}
    \label{eq:AIT}
    E_B - E_A &= \!\int_{\mathbb{R}^n} \!\! d\bm{x}\, \rho_A (\bm{x}) \,\mathcal{K}[\Delta v](\bm{x})
\end{align}
with difference in external potentials $\Delta v (\bm{x}):= v_B(\bm{x}) - v_A(\bm{x})$ and the kernel
\begin{align}
    \mathcal{K}[\Delta v](\bm{x}) := \int_0^1 d\lambda  \,\, \Delta v \left( A^{-1}(\lambda) \, (\bm{x} - \bm{b}(\lambda)) \right) \quad \text{.}
\end{align}
But how to obtain the quantities $A(\lambda)$ and $\bm{b}(\lambda)$? Consider the hydrogen-like (HL) atom from Ref.~\onlinecite{krug_generalAIT} as example: just the problem's statement, i.e. the (electronic) Hamiltonian, reads:
\begin{align}
    \label{eq:Hamiltonian_hydlike}
    \hat{H}^{\rm HL} :=& - \frac{1}{2} \bm{\nabla}^2_{\bm{x}} - \frac{Z_A}{||\bm{x}||_2} \\
    = &\, Z_A^2 \left(- \frac{1}{2} \frac{\bm{\nabla}^2_{\bm{x}}}{Z_A^2} - \frac{1}{Z_A||\bm{x}||_2} \right)
\end{align}
Simply rescaling of $\bm{x} \rightarrow (Z(\lambda)/Z_A) \, \bm{x}$ does the trick of transforming the coordinates of the Hamiltonian at nuclear charge $Z_A$ to a general one at $Z(\lambda)$:
\begin{align}
    \hat{H}^{\rm HL} \rightarrow Z_A^2 \left(- \frac{1}{2} \frac{\bm{\nabla}^2_{\bm{x}}}{Z^2(\lambda)} - \frac{1}{Z(\lambda)||\bm{x}||_2} \right)
\end{align}
Our affine transformation is just a factor $A(\lambda) = Z(\lambda)/Z_A$, which results in the kernel
\begin{align}
    \label{eq:kernel_hydlike}
    \mathcal{K}[\Delta v](\bm{x})
    &= \frac{-Z_B+Z_A}{2||\bm{x}||_2} \left(1 + \frac{Z_B}{Z_A} \right) .
\end{align}
See Ref.~\onlinecite{krug_generalAIT} for further details. 

Consider now a monoatomic, multi-electron system with fictitious inter-electron potential ($\propto$ distance$^{-2}$),
\begin{align}
    \label{eq:1/r2_Hamiltonian}
    &\hat{H}^{\rm fic} := \sum_{i} \left(- \frac{1}{2} \bm{\nabla}^2_{\bm{x}_i} - \frac{Z_A}{||\bm{x}_i||_2} + \frac{1}{2} \sum_{j\neq i} \frac{1}{||\bm{x}_i - \bm{x}_j||_2^2}\right) \\
    &= Z_A^2 \sum_{i} \left(- \frac{1}{2} \frac{\bm{\nabla}^2_{\bm{x}_i}}{Z_A^2} - \frac{1}{Z_A||\bm{x}_i||_2} + \frac{1}{2} \sum_{j\neq i} \frac{1}{Z_A^2 ||\bm{x}_i - \bm{x}_j||_2^2}\right)
\end{align}
This is rescalable by the  same transformation as in the hydrogen-like atom 
and thus produces the same kernel. 


Unfortunately, for the real multi-electron atom,
\begin{align}
    \label{eq:atomic_Hamiltonian}
    \hat{H}^{\rm atom} := \sum_{i} \left(- \frac{1}{2} \bm{\nabla}^2_{\bm{x}_i} - \frac{Z_A}{||\bm{x}_i||_2} + \frac{1}{2} \sum_{j\neq i}\frac{1}{||\bm{x}_i - \bm{x}_j||_2}\right) 
\end{align}
no scaling transformation (or any affine transformation for that matter) is available to produce a kernel as in Eq.~\ref{eq:kernel_hydlike}. 
However, since both the systems with constant (or no) electron-electron interaction (Eq.~\ref{eq:Hamiltonian_hydlike}), and inversely quadratic interaction (Eq.~\ref{eq:1/r2_Hamiltonian}) produce the same kernel, 
we now assume the hydrogen-like kernel of Eq.~\ref{eq:kernel_hydlike} to correspond to a fair approximation.

$\hat{H}^{\rm fic}$ in Eq.~\ref{eq:1/r2_Hamiltonian} highlights a limitation of applying AIT for multi-electron systems: both the kinetic terms and the inter-electron repulsion include no transformable parameter (like the nuclear charge). Thus, both of them need the same dimension in the coordinates, namely distance$^{-2}$, for them to remain consistent with one another when employing the parametrization~$\bm{r}(\lambda)$. Real electrons, however, experience a repulsion proportional to distance$^{-1}$.

We considered AIT for monoatomics previously in Ref.~\onlinecite{krug_DeltaE}. There, however, we found a parametrization by trial and error. Although this produced reasonable results, it left us without any measure to assess its error and systematically improve upon it. In this paper, it is quite clear from comparison of the problem statement in Eq.~\ref{eq:atomic_Hamiltonian} to the Hamiltonian in Eq.~\ref{eq:1/r2_Hamiltonian} that our error introduced by choosing the (approximate) transformation $\bm{x} \rightarrow Z(\lambda)/Z_A \, \bm{x}$ scales with the ratio $Z_B/Z_A$ because the strength of inter-electron repulsion is miss-scaled by $Z(\lambda)/Z_A$.

Using this approximation with Eq.~\ref{eq:AIT}, 
we can now write the energy difference between iso-electronic atoms~$A$ and~$B$ approximately as:
\begin{align}
    \label{eq:methods_formula_derived}
    E_B - E_A \approx \frac{-Z_B^2 + Z_A^2}{2Z_A} \underbrace{\int_{\mathbb{R}^3} \!\! d\bm{x}\, \frac{\rho_A(\bm{x})}{||\bm{x}||_2}}_{=: \mu_A}
\end{align}
The integral is known as the ('alchemical') electrostatic potential at the nucleus,~$\mu_A$. Eq.~\ref{eq:methods_formula_derived} looks very similar to Levy's formula for energy differences from averaged electron densities when applied to the case of iso-electronic
atoms\cite{levy_1978,levy2008}. 
Here, however, we only rely on knowledge about $\rho_A$, not an average electron density!

\begin{figure}
    \centering
    \includegraphics[width=\linewidth]{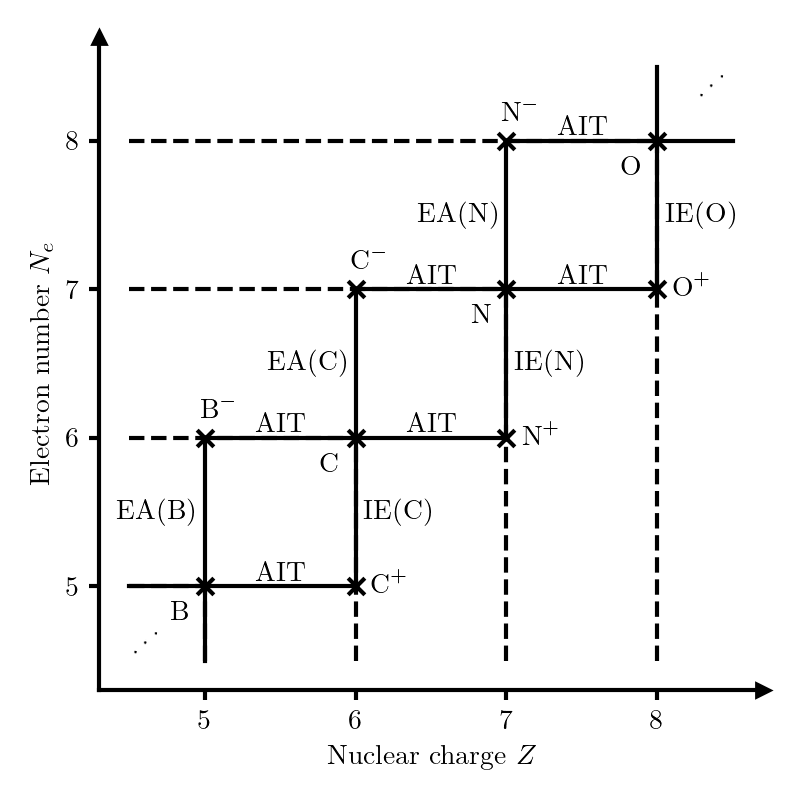}
    \caption{Overview of changes for single atoms at different nuclear charges $Z$ and electron numbers $N_e$. Horizontal and vertical lines denote iso-electronic alchemical changes (AIT), corresponding to addition or removal of protons from the nucleus, and electron number changes, i.e. electron affinity (EA) and ionization energy (IE), respectively.}
    \label{fig:cartoon_cycles}
\end{figure}

Eq.~\ref{eq:methods_formula_derived} allows estimates of iso-electronic energy trends in atomic ions, and thus tackles the missing link for navigating atoms of any nuclear charge $Z$ and electron number $N_e$\cite{vonlilienfeld2023_1st_order}. The iso-protonic analogon, i.e. electron affinities (EA) and ionization energies (IE) for fixed nuclear charges, are experimentally measured quantities in physical chemistry. Combining both changes into thermodynamic cycles, one can arrange all elements and their possible ions in a scheme as given in Fig.~\ref{fig:cartoon_cycles}.

\section{Results and Discussion}
\label{sec:results}

\begin{figure*}
    \centering
    \includegraphics[width=\linewidth]{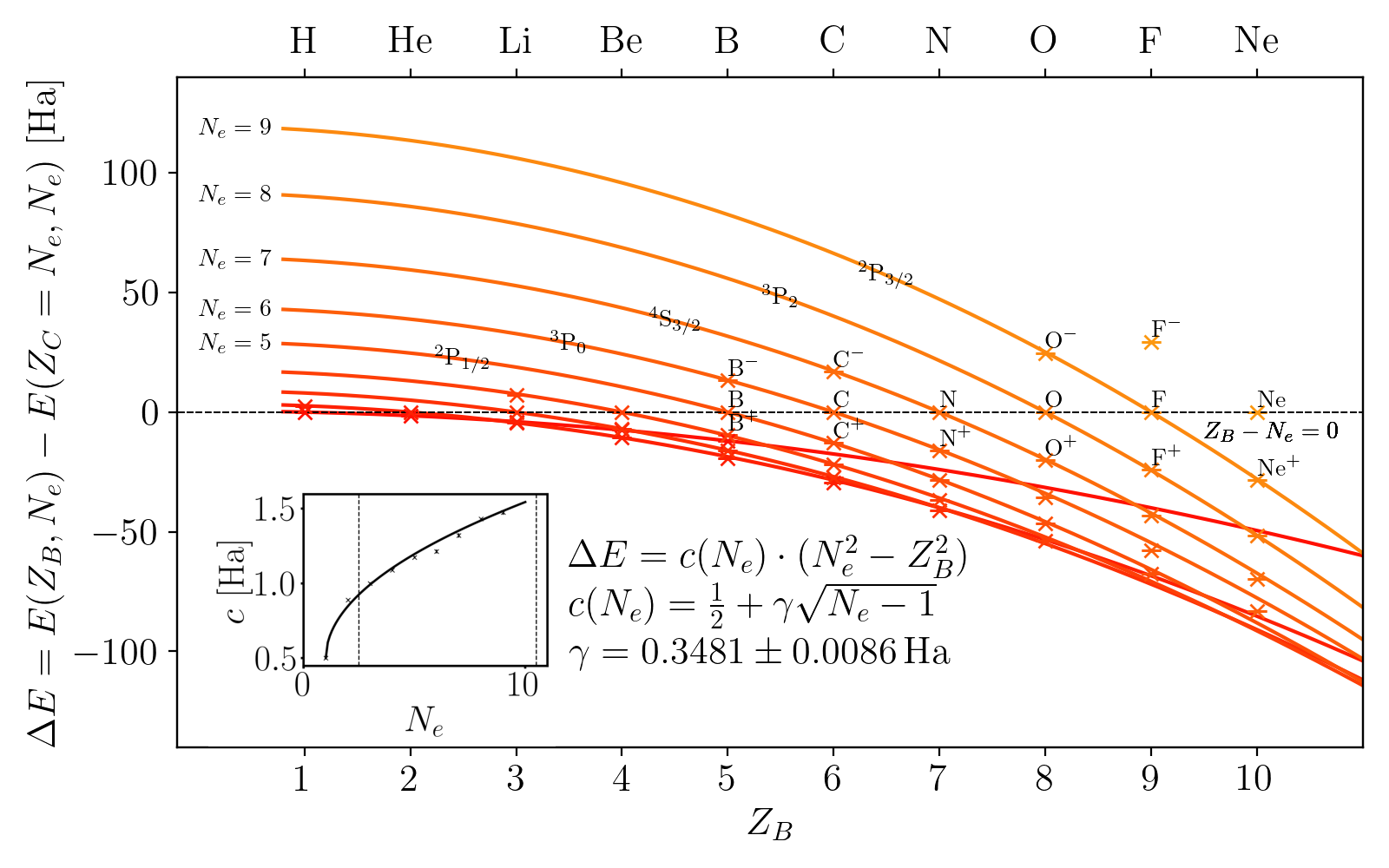}
    \caption{Experimental energy differences $\Delta E$ vs. the final system's nuclear charge $Z_B$ for different iso-electronic atoms $Z_B, Z_{C} \in \lbrace 1, \dots, 10 \rbrace$ if initial and final atom's electronic configuration match, electron number $N_e \in \lbrace Z_B+2, \dots , Z_B-4 \rbrace$ and constraint $Z_{C} = N_e$. Iso-electronic fits are solid, colored lines. Term symbols of the neutral reference are given along the fitted lines. Inset: Fitted values for the parameter $c$ vs. $N_e$ for each iso-electronic series with fit function.}
    \label{fig:DeltaE_vs_Z_exp_small}
\end{figure*}

In the methods section, we derived a formula for the approximate, non-relativistic energy difference between two iso-electronic atoms $A$ and $B$:
\begin{align}
    E_B - E_A \approx \left(Z_A^2 - Z_B^2 \right) \frac{\mu_A}{2Z_A} \quad \text{,}
\end{align}
where $\mu_A$ denotes the ('alchemical') electrostatic potential at the nucleus of $A$~\cite{levy_politzer1987,lilienfeld_variational, anatole2006moleculargrandcanonicalensemble}. 
Considering the relative energy difference of $A$ to any third iso-electronic atom $C$, the $E_A$ contribution cancels and the remaining energy difference between atom $B$ and $C$ reads:
\begin{align}
    \label{eq:B_minus_C}
    \Delta E := E_B - E_C \approx (Z_C^2 -Z_B^2) \frac{\mu_A}{2Z_A}
\end{align}
Note that for $Z_C - Z_A = Z_A - Z_B$, this equation recovers the energy difference exactly up to 4th order in an alchemical Taylor expansion~\cite{vonlilienfeld2023_1st_order}.
However, since system $A$ is independent of $Z_B, Z_{C}$, the ratio $\mu/Z$ must be a constant --- for any fixed electron number $N_e$.

Such a formula for energy differences is not new: in 1978, Levy found\cite{levy_1978}:
\begin{align}
    \label{eq:levy_1978}
    E_B - E_C &= \int_{Z_C}^{Z_B} \! dZ \int_{\mathbb{R}^3} \!\! d\bm{x}\, \frac{\rho_Z(\bm{x})}{||\bm{x}||_2} \\
    \label{eq:levy_1978_meanvalue}
    &\approx (Z_B - Z_C) \int_{\mathbb{R}^3} \!\! d\bm{x}\, \frac{\bar{\rho}(\bm{x})}{||\bm{x}||_2}
\end{align}
In the second line, he employed the mean-value theorem and $\bar{\rho} = (\rho_C + \rho_B)/2$.

Both his and our observations are consistent with the theoretical and numerical findings by Levy, Tal and Clement that the integral in~$\mu$ "appears to be remarkably close to a linear function of $Z$"\cite{levy_1982}, which was later corroborated by Politzer and Levy using experimental data\cite{levy_politzer1987}. This linearity is a necessity for Eq.~\ref{eq:B_minus_C} to be (approximately) independent of $Z_A$ \footnote{We encounter no discontinuities as discussed in detail in their paper \cite{levy_1982} because we force $\rho_A$ to be iso-electronic. Since electrons cannot leave system $A$ here, even if energetically favorable, all functional relations are smooth} and the integral in Eq.~\ref{eq:levy_1978_meanvalue} to be proportional to $Z_B+ Z_C$ which implies the quadratic behavior of $\Delta E$ in $Z$.

However, already before these findings, an exact formula for non-relativistic ground-state energy (not just energy differences!) was provided by Politzer and Parr in 1974\cite{Politzer_Parr_1974}, substantiating the quadratic $Z$-behavior of single atoms. In this regard, AIT can be considered quite late to the party. 
However, all of these formulas make approximations to the electron density to obtain predictions, either via the mean-value theorem as in Eq.~\ref{eq:levy_1978}, or by pre-computing the "free-atom screening function"\cite{Politzer_Parr_1974}.
AIT, in contrast, makes its approximations already in the transform used to connect the Hamiltonians. This is a conceptual difference: the methods mentioned above establish approximations to the electron density from an electrostatic viewpoint, while AIT does so to the underlying Schrödinger equation from a mathematical one. The former requires additional knowledge about the quantum nature of the density to justify the approximation, the latter gives a straight-forward estimate for its error.

As a check of validity, we are quite satisfied with reproducing the earlier and more established formulas, while taking a new path to do so, hence revealing new perspectives.

Eq.~\ref{eq:B_minus_C} gives not just access to the total energy between two atoms; by virtue of the virial theorem \cite{cohentannoudji} for single atoms with electrons in the Coulomb potential, $<T> = -<V>/2$, one immediately finds both the kinetic and potential energy contributions as well.

\subsection{Comparison to experimental data for Hydrogen to Radon}

Varying only $Z_B$ and fixing $Z_C$ to correspond to the charge-neutral atom (i.e.~$N_e = Z_C$), Fig.~\ref{fig:DeltaE_vs_Z_exp_small} displays the experimental $\Delta E$ values\cite{NIST,electron_affinities_rienstra-kiracofe} approximately quadratic in $Z_B$. 
In the case of just one electron, i.e.~hydrogen-like series, we must recover $c = \mu_C/2Z_C = 1/2$. With this constraint, we find that least square regression yields good agreement for the following form,
\begin{align}
    \label{eq:c_explicitly}
    c(N_e) = \frac{1}{2} + \gamma \sqrt{N_e - 1},
\end{align}
for $N_e \ge 1$ (see inset of Fig.~\ref{fig:DeltaE_vs_Z_exp_small}).
Here, $\gamma$ is assumed to be one universal parameter for all atoms and ions, independent of electron number.
Combining Eqs.~\ref{eq:B_minus_C} and~\ref{eq:c_explicitly}, and using experimental energies for all neutral atoms and ions with no more than 10 electrons and protons, we fit $\gamma = 0.3481 \pm 0.0086$~Ha to reproduce $\Delta E$ in terms of number of electrons~$N_e$. 
Fitting to the experimental data for all atoms up to Radon yields
$\gamma = 0.3766 \pm 0.0020$~Ha. 
Corresponding figures with all elements up to Radon can be found in the Supplemental Material (Fig.~\ref{fig:DeltaE_vs_Z_exp_conservedlevels}). 
Care has been taken, as AIT only applies for a given electronic state, i.e.~if the electronic states of initial and final systems match. 
A version without this constraint, together with an analogue figure with data from DFT instead of experiment, can be found in the Supplemental Material as well (Figs.~\ref{fig:DeltaE_vs_Z_exp_conservedlevels},\ref{fig:DeltaE_vs_Z_exp},\ref{fig:DeltaE_vs_N_exp},\ref{fig:DeltaE_vs_Z_DFT}).
Throughout all data, we find correlation coefficients of $R^2 > 0.993$.

Suggesting a functional form of $c$, i.e. Eq.~\ref{eq:c_explicitly}, we assert generality but loose accuracy. Any fit of $\gamma$ imposes a functional form on the electrostatic potential $\mu$ but without theoretical justification or inclusion of relativistic effects, the latter being known to introduce energy corrections $\propto Z^4$ (cf.~fine structure\cite{cohentannoudji}). If instead one fits~$c$ for each electron number anew, one evades this at the cost of generality.
We want to emphasize the strictly non-relativistic approach of AIT for the Hamiltonians in Eqs.~\ref{eq:Hamiltonian_hydlike}, \ref{eq:1/r2_Hamiltonian} and~\ref{eq:atomic_Hamiltonian} in specific, although the electron density needed for predictions may be obtained with inclusion of relativistic effects. 

Originally, we expected Eq.~\ref{eq:c_explicitly} to be proportional to $N_e^{7/3}$ as this is the functional relation of the total binding energy of the neutral atom in the Thomas-Fermi model\cite{Englert_book_1988} (which, unfortunately, holds exactly only in the limit of infinite nuclear charge). However, an exponent of~$1/2$ yields a better fit to the experimental data.

Reinsertion into Eq.~\ref{eq:B_minus_C} provides a general approximate formula for energy differences between atoms in any fixed iso-electronic series,
\begin{align}
    \label{eq:DeltaE_atoms}
    \Delta E &\approx \left( \frac{1}{2} + \gamma \sqrt{N_e - 1} \right) (Z_C^2 - Z_B^2) \\
    & = -\left( 1 + 2\gamma \sqrt{N_e - 1} \right) \, \Delta Z \, \Bar{Z}
\end{align}
with difference in nuclear charge $\Delta Z := Z_B - Z_C$ and their mean $\Bar{Z} := (Z_B+ Z_C)/2$.

\subsection{The He-like series}

In general, one may consider any one iso-electronic series in particular, then pick any trust-worthy experimental energy difference inside this series to calibrate $c$ (without constraint to the functional form of Eq.~\ref{eq:c_explicitly}), and subsequently apply Eq.~\ref{eq:B_minus_C} to obtain all the remaining values. 
To assess our formula's predictive power, we have chosen the He-like atoms for their historical significance \cite{hylleraas_1956,baker_freund}, and for the availability
of experimental, theoretical (via perturbation theory), and computational data.
\begin{figure}
    \centering
    \includegraphics[width=\linewidth]{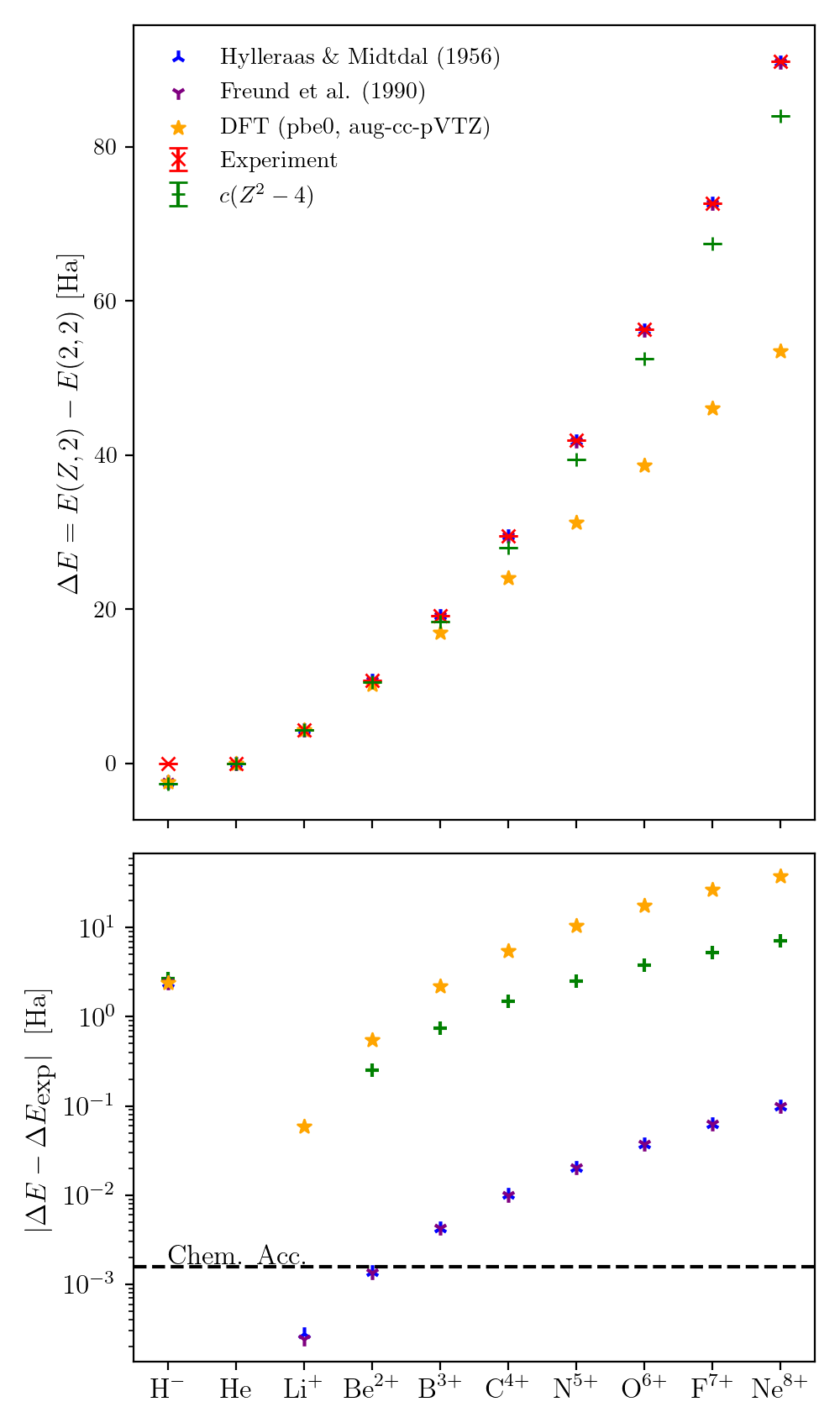}
    \caption{Performance of our formula (green symbols, $c =$\,0.8752 Ha). 
    Predicted energy differences of He-like ions relative to neutral He from different sources (top) as a function of nuclear charge. 
    For comparison, literature values are given for experiments~\cite{NIST,electron_affinities_rienstra-kiracofe}, perturbation theory by Hylleraas and Midtdal~\cite{hylleraas_1956}, perturbation theory by Freund et al.~\cite{baker_freund}, and DFT calculations (this work). 
    Absolute error with respect to experimental data (bottom). }
    \label{fig:hylleraas_vs_exp_vs_AIT}
\end{figure}

Using just the energy difference between Li$^+$ and He\cite{NIST,electron_affinities_rienstra-kiracofe} we determine $c$ to correspond to 0.8752~Ha with the experimental uncertainty being less than 10$^{-8}$. 
Predictions and unsigned errors for the remainder of the entire He-like series are shown in Fig.~\ref{fig:hylleraas_vs_exp_vs_AIT}, along with measurements and results from other methods. 

Note how DFT (using the exchange-correlation functional \texttt{pbe0}\cite{PBE0,PBE01,PBE02} and basis set \texttt{aug-cc-pVTZ}\cite{aug-cc-pVTZ_1,aug-cc-pVTZ_2,aug-cc-pVTZ_3,aug-cc-pVTZ_4,aug-cc-pVTZ_5,aug-cc-pVTZ_6}), is outperformed by our {\em Bierdeckel} estimate. 
We also compare to Hylleraas' historic perturbative $1/Z$-expansion using five terms \cite{hylleraas_1956}, as well as to a subsequent contribution relying on over 500 terms~\cite{baker_freund}, both of which still fall short of chemical accuracy w.r.t. the experimental values \cite{NIST, electron_affinities_rienstra-kiracofe} for all ions but Li$^+$ and Be$^{2+}$.

We are fully aware of the plethora of exchange-correlation-functionals\cite{tsuneda_2014}, the different types of computational methods to solve for He-like energies aside DFT (coupled cluster, full contact interaction, etc.), and their merit in the diverse and complex applications beyond calculations of single atoms. The data in Fig.~\ref{fig:hylleraas_vs_exp_vs_AIT} serves not to undermine these methods, but to classify Eq.~\ref{eq:DeltaE_atoms} in relationship to a widely used, general purpose method (DFT), with a widely used functional (\texttt{pbe0}), to highlight the accuracy of alchemical methods as a general, and very inexpensive, approach to the problem of energy predictions. 

\subsection{Prediction of electron affinities from ionization energies}


Electron affinities can be difficult to measure. 
The above established connection between atoms of one iso-electronic series suggest that the energies of neutral atoms $Z,Z+1$ in conjunction with AIT (i.e. parameters $c$ and thus ionization energies) might suffice to predict the (first) electron affinities of atoms $Z$. Subtracting Eq.~\ref{eq:B_minus_C} for $Z,Z+1$ and $N_e,N_e+1$, we find:
\begin{align}
    \label{eq:EA_pred_formula}
    EA(Z) := & \, E(Z,N_e+1) - E(Z,N_e)\notag \\
    = & \, (2Z+1)[c(N_e+1) - c(N_e)] \notag \\
    & + E(Z+1,N_e+1) - E(Z+1,N_e)
\end{align}
For this, AIT requires initial and final state to be identical which renders direct comparison to experiments problematic, as in experiment, states often change between atoms $Z$ and $Z+1$ (or iso-electronic ions) and consequently, AIT is not applicable.

In addition, Eq.~\ref{eq:EA_pred_formula} suffers from unfavorable error propagation: subtracting two large numbers, $E(Z+1,N_e+1) - E(Z+1,N_e)$, to obtain a small one, in addition to the scaling of $c$'s error with nuclear charge~$Z$, carry both significant errors and render a sound prediction of electron affinities unfeasible. A visualization of this can be found in the Supplemental Material (Fig.~\ref{fig:EA_pred_exp}).

\section{Conclusion}
\label{sec:conclusion}

We motivated and derived a kernel of AIT for monoatomic systems. This led to an approximate proportionality of the relative energy of any two iso-electronic atoms being purely quadratic in their nuclear charges, together with a general formula for the energy difference of atoms in Eq.~\ref{eq:DeltaE_atoms}. We numerically tested this approach with experimental data for ionization energies and electron affinities of the entire periodic table, together with additional tests using DFT data. The He-like energies were treated in detail.

The utility of Eq.~\ref{eq:DeltaE_atoms} extends beyond single atoms, as many methods applied in molecules, e.g. for chemical reactions, bonding energies and distances, are intimately related to the energy of their constituent atoms as recently discussed from the alchemical viewpoint \cite{michael_hammett_JACS}. 


Future work will deal with kernels beyond single atoms, e.g. in constitutional isomers or as approximate treatments of molecules. Both could be rendered arbitrarily accurate when used as baseline models for $\Delta$-ML, multi-fidelity ML, or within transfer learning.
In addition, non-relativistic corrections from the fine structure could be considered.

Beyond these direct applications, we note that since the advent of the periodic table, arranging systems by their nuclear charge proved sensible\cite{pettifor_1984}. Not just from the point of quantum alchemy, but computational chemistry as well, classifying systems by their electron number might be a fruitful concept, as proven by the content of this study: neither did we consider new data, nor new computational methods; the alchemical perspective alone revealed new and simple relationships.

\section*{Supplemental Material}
\begin{sidewaysfigure}
    \includegraphics[width=\textwidth]{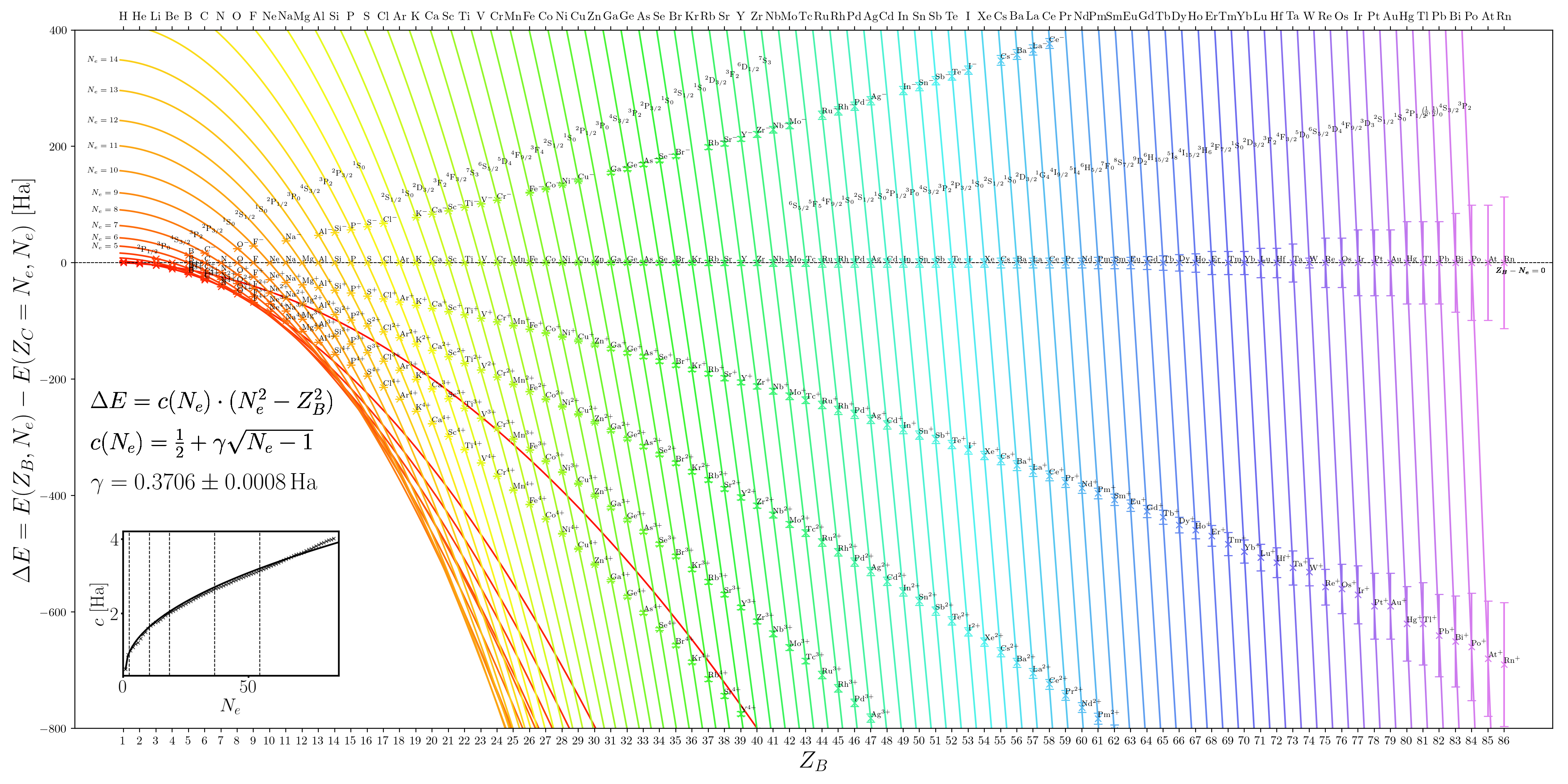}
    \caption{Experimental energy differences $\Delta E$ vs. the final system's nuclear charge $Z_B$ for different iso-electronic atoms $Z_B, Z_{C} \in \lbrace 1, \dots, 86 \rbrace$, without the constraint of matching initial and final atom's electronic configuration, electron number $N_e \in \lbrace Z_B+2, \dots Z_B-4 \rbrace$ and $Z_{C} = N_e$. Iso-electronic fits are solid, colored lines. Term symbols of the neutral reference are given along their respective fitted lines. Inset: Fitted values for the parameter $c$ vs. $N_e$ for each iso-electronic series.}
    \label{fig:DeltaE_vs_Z_exp}
\end{sidewaysfigure}

\begin{sidewaysfigure}
    \includegraphics[width=\textwidth]{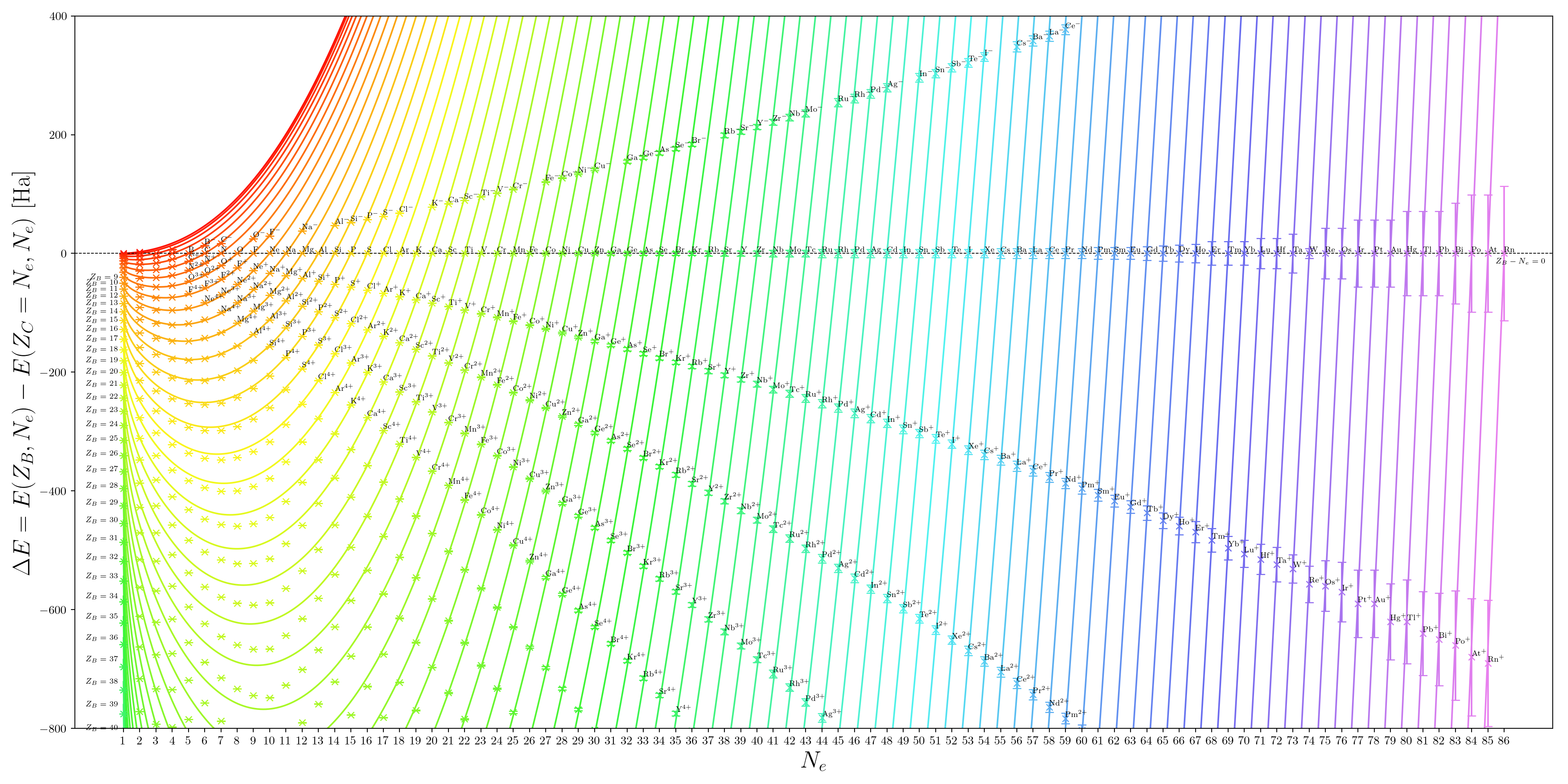}
    \caption{Experimental energy differences $\Delta E$ vs. the electron number $N_e$ for different atoms $Z_B, Z_{C} \in \lbrace 1, \dots, 86 \rbrace$, without the constraint of matching initial and final atom's electronic configuration and $Z_{C} = N_e$. Iso-atomic fits are solid, colored lines. The fit is identical to the one in Fig.~\ref{fig:DeltaE_vs_Z_exp}. Even more extreme cations than +4 are drawn but never used for the fit and only serve orientation and comparison.}
    \label{fig:DeltaE_vs_N_exp}
\end{sidewaysfigure}

Versions of Fig.~\ref{fig:DeltaE_vs_Z_exp_small} with all elements up to Radon (Fig.~\ref{fig:DeltaE_vs_Z_exp_conservedlevels}) or without the constraint of matching electronic states (Fig.~\ref{fig:DeltaE_vs_Z_exp}) or with DFT numbers instead of experimental numbers (Fig.~\ref{fig:DeltaE_vs_Z_DFT}) can be found in the Supplemental Material, together with Fig.~\ref{fig:EA_pred_exp}.

\section*{Data and code availability}

The code that produces the figures and findings of this study, in specific the scripts for the generation of DFT data, plotting and fitting, are openly available on Zenodo under \url{zenodo.org/records/12547814}. The ionization energies were obtained from the National Institute of Standards and Technology \cite{NIST}, the electron affinities were taken from the review by Rienstra-Kiracofe et al. \cite{electron_affinities_rienstra-kiracofe}. Both datasets are also accessible on Zenodo.

\section*{Software}
Software for the purpose of data generation (e.g. quantum chemistry software) are provided by the \texttt{Python}-packages \texttt{PySCF} \cite{PySCF1,PySCF2}, \texttt{basissetexchange} \cite{bse}, \texttt{NumPy} \cite{numpy} and \texttt{SciPy} \cite{scipy}. Visualizations were created using \texttt{Matplotlib} \cite{matplotlib}.

\section*{Acknowledgements}
We acknowledge discussions with Kieron Burke, Roi Baer, Dirk Andrae, Florian Bley and Danish Khan.
We acknowledge the support of the Natural Sciences and Engineering Research Council of Canada (NSERC), [funding reference number RGPIN-2023-04853]. Cette recherche a été financée par le Conseil de recherches en sciences naturelles et en génie du Canada (CRSNG), [numéro de référence RGPIN-2023-04853].
This research was undertaken thanks in part to funding provided to the University of Toronto's Acceleration Consortium from the Canada First Research Excellence Fund,
grant number: CFREF-2022-00042.
O.A.v.L. has received support as the Ed Clark Chair of Advanced Materials and as a Canada CIFAR AI Chair.
O.A.v.L. has received funding from the European Research Council (ERC) under the European Union’s Horizon 2020 research and innovation programme (grant agreement No. 772834).

\section*{Author Contributions}
\textbf{Simon León Krug}: conceptualization (equal), data curation, formal analysis (lead), methodology (lead), software, visualization (equal), writing - original draft (lead), writing - review \& editing (supporting).
\textbf{O.~Anatole von~Lilienfeld}: conceptualization (equal), formal analysis (supporting), methodology (supporting), funding acquisition, project administration, resources, supervision (lead), visualization (equal), writing - review \& editing (lead).

All authors read and approved the final manuscript.

\section*{Conflict of Interest}
The authors have no conflicts to disclose.


\bibliography{refs.bib}{}
\bibliographystyle{unsrt}

\clearpage
\onecolumngrid
\setcounter{section}{0}

\begin{center}
    \large \textbf{\papertitle}\\
    \large \textbf{ ---  Supplemental Information  ---}\\
    \vspace{\baselineskip}
    \normalsize Simon León Krug,$^{1}$ and O. Anatole von Lilienfeld$^{1,2,3,4,5,6,7}$\\
    \vspace{0.5\baselineskip}
    \small \textit{
    $^{1)}$Machine Learning Group, Technische Universität Berlin,
    10587 Berlin, Germany\\
    $^{2)}$Berlin Institute for the Foundations of Learning and Data, 10587 Berlin, Germany\\
    $^{3)}$Chemical Physics Theory Group, Department of Chemistry, University of Toronto, St. George Campus, Toronto, ON, Canada\\
    $^{4)}$Department of Materials Science and Engineering, University of Toronto, St. George Campus, Toronto, ON, Canada\\
    $^{5)}$Vector Institute for Artificial Intelligence, Toronto, ON, Canada\\
    $^{6)}$Department of Physics, University of Toronto, St. George Campus, Toronto, ON, Canada\\
    $^{7)}$Acceleration Consortium, University of Toronto, Toronto, ON, Canada\\
    }
    \small (Dated: \today)
\end{center}

\clearpage


\begin{sidewaysfigure}
    \includegraphics[width=\textwidth]{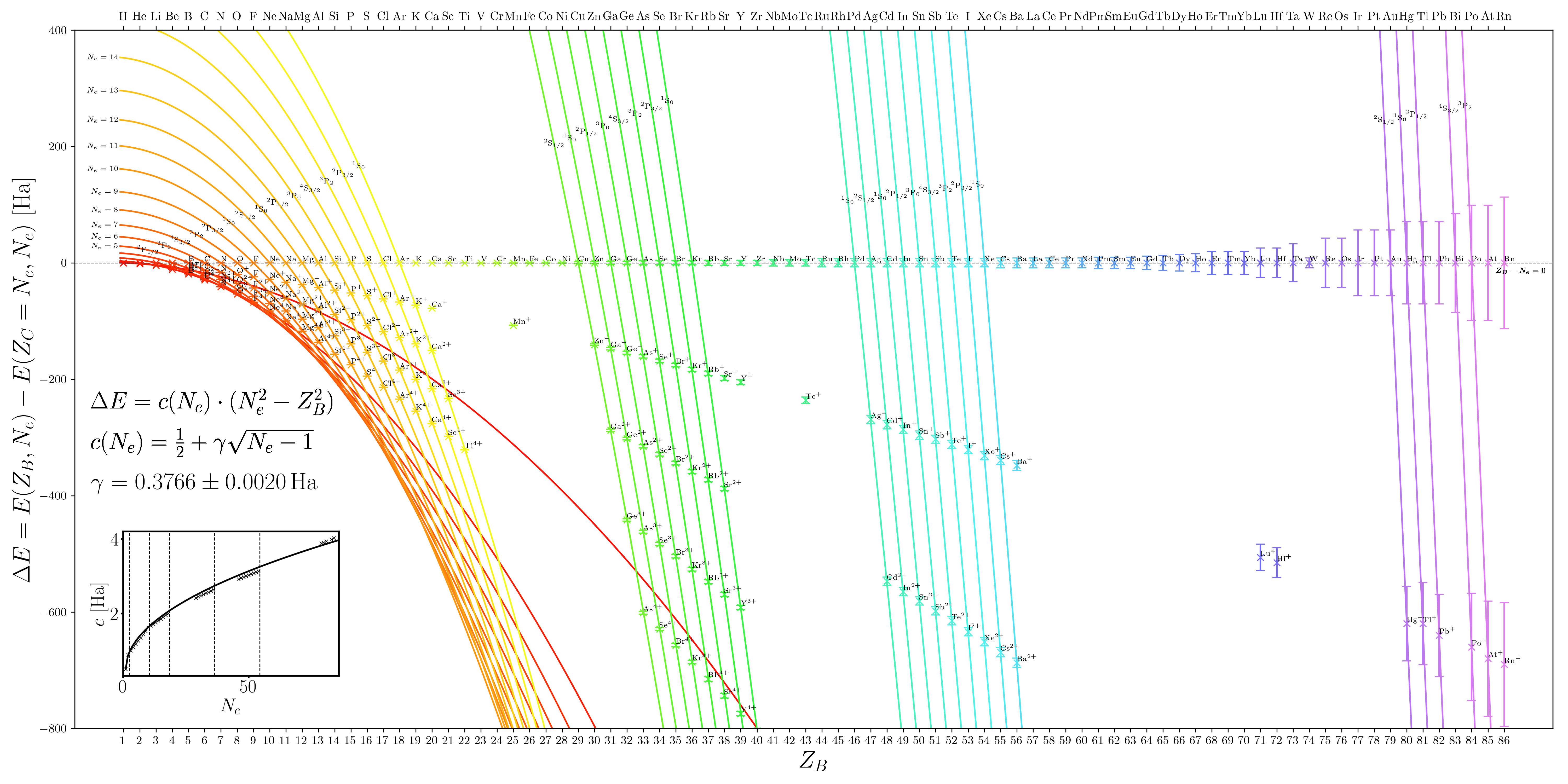}
    \caption{Experimental energy differences $\Delta E$ vs. the final system's nuclear charge $Z_B$ for different iso-electronic atoms $Z_B, Z_{C} \in \lbrace 1, \dots, 86 \rbrace$ if initial and final atom's electronic configuration match, electron number $N_e \in \lbrace Z_B+2, \dots Z_B-4 \rbrace$ and constraint $Z_{C} = N_e$. Iso-electronic fits are solid, colored lines. Term symbols of the neutral reference are given along the fitted lines. Inset: Fitted values for the parameter $c$ vs. $N_e$ for each iso-electronic series with fit function.}
    \label{fig:DeltaE_vs_Z_exp_conservedlevels}
\end{sidewaysfigure}

\begin{sidewaysfigure}
    \includegraphics[width=\textwidth]{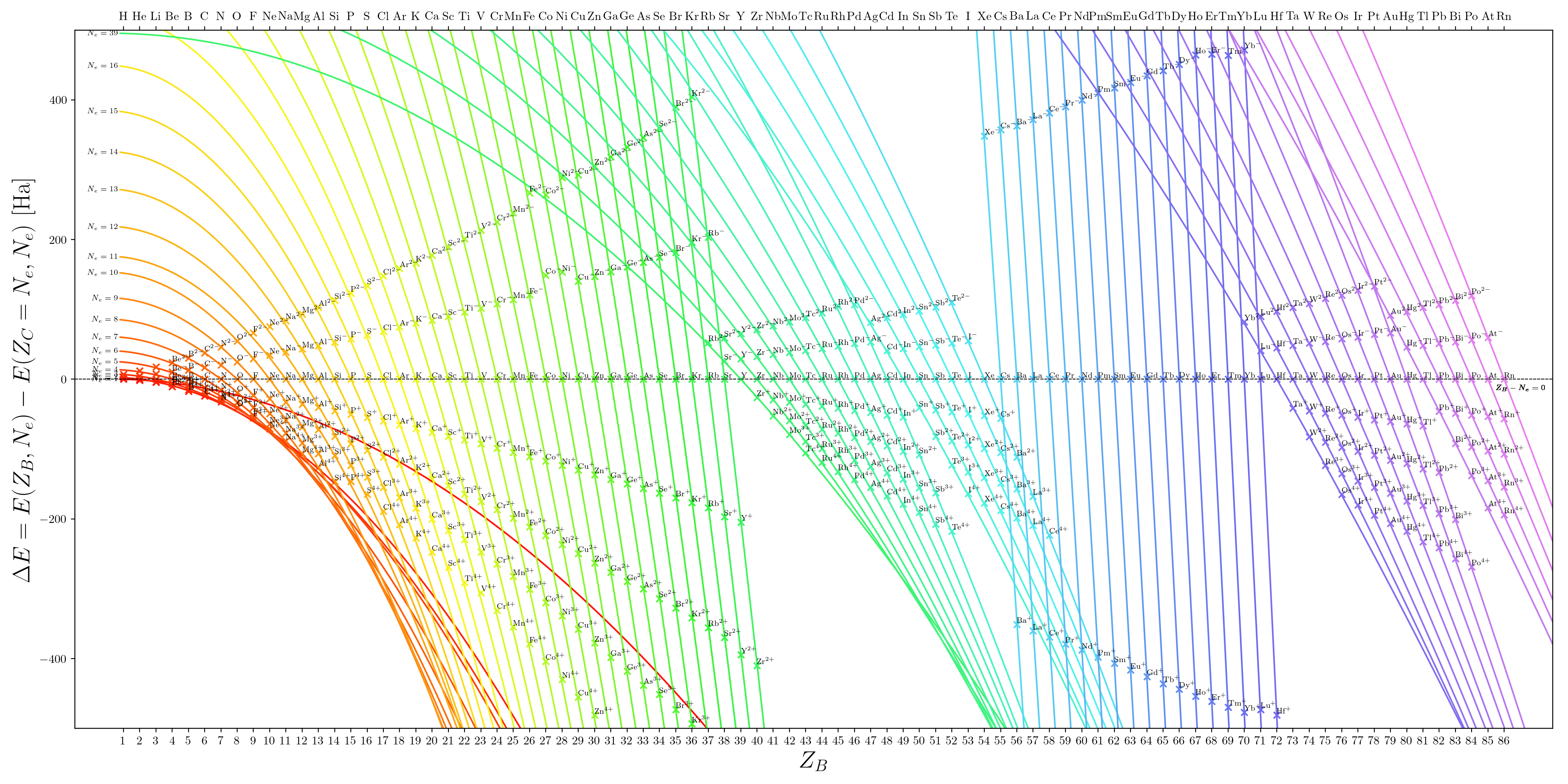}
    \caption{Energy differences $\Delta E$ from Density Functional Theory vs. the final system's nuclear charge $Z_B$ for different iso-electronic atoms $Z_B, Z_{C} \in \lbrace 1, \dots, 86 \rbrace$, electron number $N_e \in \lbrace Z_B+2, \dots Z_B-4 \rbrace$ and $Z_{C} = N_e$. Iso-electronic fits are solid, colored lines. The DFT computations were performed using the functional \texttt{pbe0}\cite{PBE0,PBE01,PBE02} and basis sets \texttt{aug-cc-pVTZ}\cite{aug-cc-pVTZ_1,aug-cc-pVTZ_2,aug-cc-pVTZ_3,aug-cc-pVTZ_4,aug-cc-pVTZ_5,aug-cc-pVTZ_6} if $N_e \in \lbrace 1, \dots, 18, 21, \dots 36 \rbrace$, \texttt{aug-cc-pVTZ-X2C}\cite{aug-cc-pVTZ-X2C} if $N_e \in \lbrace 19,20,37,38,55,56 \rbrace$, \texttt{aug-cc-pVTZ-PP}\cite{aug-cc-pVTZ-PP_1,aug-cc-pVTZ-PP_2,aug-cc-pVTZ-PP_3,aug-cc-pVTZ-PP_4,aug-cc-pVTZ-PP_5} if $N_e \in \lbrace 39, \dots, 54, 72, \dots 86 \rbrace$ and \texttt{Sapporo-DKH3-TZP-2012-diffuse}\cite{Sapporo} if $N_e \in \lbrace 57, \dots 71 \rbrace$. For full comparability between the ions, each ion's spin was fixed to $N_e\!\! \mod 2$.}
    \label{fig:DeltaE_vs_Z_DFT}
\end{sidewaysfigure}

\begin{figure}
    \centering
    \includegraphics[width=0.8\textwidth]{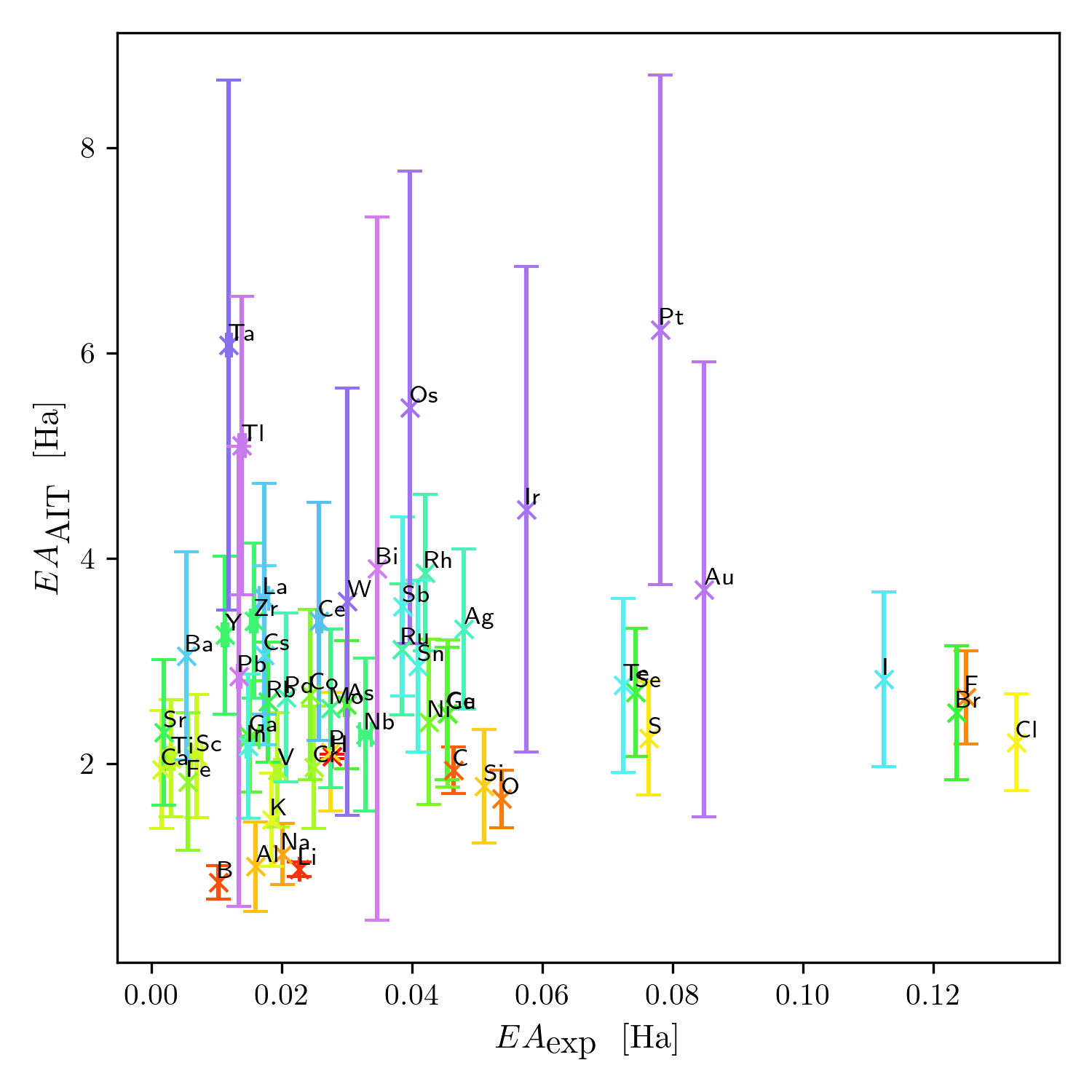}
    \caption{Prediction of electron affinities from experimental data \cite{NIST} vs. from AIT without the constraint of matching term symbols (quantum numbers).}
    \label{fig:EA_pred_exp}
\end{figure}

\end{document}